\documentclass[aps,prstab,reprint,groupedaddress,showpacs,floatfix]{revtex4-1}

\usepackage{amsmath}
\usepackage{bm}
\usepackage[caption=false,lofdepth,lotdepth]{subfig}
\usepackage{graphicx}

\begin{document}


\title{High Power Fibre Laser System for a High Repetition Rate Laserwire}


\author{L.~J.~Nevay}
\email[]{laurie.nevay@rhul.ac.uk}
\altaffiliation{now at Royal Holloway, University of London}
\author{R.~Walczak}
\author{L.~Corner}
\affiliation{John Adams Institute at the University of Oxford, Denys Wilkinson Building, Keble Road, Oxford, OX1 3RH United Kingdom}


\date{\today}

\begin{abstract}
We present the development of a high power fibre laser system to investigate its suitability 
for use in a transverse electron beam profile monitor i.e. a laserwire. A system capable 
of producing individual pulses up to 165.8~$\pm$~0.4~$\mu$J at 1036~nm with a full-width at
half-maximum of 1.92~$\pm$~0.12~ps at 6.49~MHz 
is demonstrated using a master oscillator power amplifier design with a final amplification 
stage in a rod-type photonic crystal fibre. The pulses are produced in trains of 1~ms
 in a novel burst mode amplification scheme to match the bunch pattern of the 
charged particles in an accelerator. 
This method allows pulse energies up to an order of magnitude greater than the steady-state
value of 17.0~$\pm$~0.6~$\mu$J to be achieved at the beginning of the burst with a demonstrated 
peak power of 25.8~$\pm$~1.7~MW after compression. The 
system is also shown to demonstrate excellent spatial quality with an $M^{2}$~=~1.26~$\pm$~0.01
 in both dimensions which would allow nearly diffraction limited focussing to be achieved. 
\end{abstract}

\pacs{07.60.Vg, 42.60.By, 29.27.Fh}

\maketitle


\section{Introduction\label{sec:introduction}}


For a future linear electron-positron collider such as the International Linear Collider
 (ILC)~\cite{ilc2013} and the Compact Linear Collider (CLIC)~\cite{clic2012}, the generation
and transport of beams while preserving the low emittance is essential to achieve
the required final focus beam sizes and therefore high luminosity. This requires
measurement, monitoring and control of the transverse beam emittance along the 
entire accelerator chain, preferably in a non-invasive manner.
In a linear accelerator, one possible method of measuring 
the emittance is by measuring the transverse size of the beam at several points in the lattice 
with different betatron phases. Two methods to achieve this are wire-scanners 
that measure a projection of the beam and optical transition 
radiation (OTR) screens to image the beam directly~\cite{Brandt2009}, 
but these can both suffer damage from high charge density beams. Conventional 
methods such as these are disruptive to the beam and they
cannot be used for measurement and tuning of the accelerator during operation,
therefore the development of new diagnostics is essential. 

\begin{figure}
\includegraphics[width=0.48\textwidth]{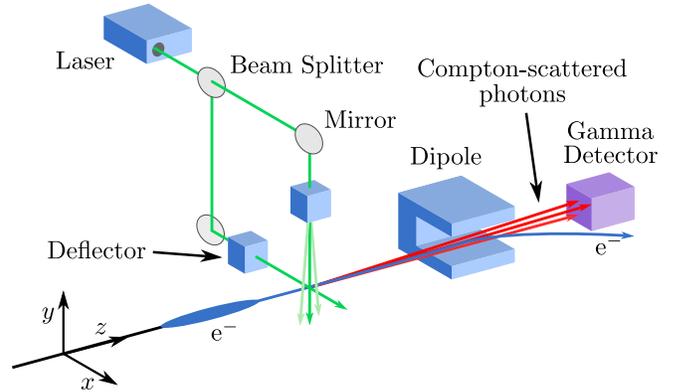}
\caption{\label{fig:alaserwire} Conceptual schematic of a laserwire. A laser beam is
scanned perpendicularly to an electron beam producing Compton-scattered photons that
are detected after a dipole magnet.}
\end{figure}

A laserwire is a beam size diagnostic in which a tightly focussed laser beam is scanned 
perpendicularly across an electron beam~\cite{Tenenbaum1999}. A small fraction of the 
laser photons undergo Compton-scattering, resulting in electrons with degraded energies 
and high energy photons that travel nearly parallel to the electron beam. A dipole magnet 
further along the accelerator separates these from the primary
electron beam as shown in Figure~\ref{fig:alaserwire}. 
The Compton-scattered photons can be detected further along the beam line after
a bending magnet where the photons and electrons are separated. 
Alternatively, in the case of a high energy collider \mbox{($E_{e}\,>\,50$~GeV)}, 
the energy of the electrons is degraded by a large fraction and it would be 
possible to detect these electrons as they are lost from the lattice.
Both can be used as a measure of the Compton-scattering rate, which is modulated
as the laser focus is scanned transversely across the electron beam,  
providing a laserwire scan. With knowledge of the laser size at its focus, the scan can be 
deconvolved to yield the electron beam size~\cite{Agapov2007}.

For accurate determination of the electron beam size, the laser spot size must be 
of a similar or ideally smaller size than the electron beam. The laser must 
be of sufficient intensity to produce a 
detectable number of Compton-scattered photons relative to the 
detector background environment. Furthermore, it must have the excellent 
spatial beam quality required to create the focussed spot sizes of approximately a
micrometre that 
will be required for use in a laserwire, at the ILC or CLIC. If a pulsed 
laser source is used, it must also have pulses of a similar duration to that of the 
individual electron bunches for optimal laser-electron collision luminosity, which 
would for example imply a pulse duration of approximately 1~ps for the ILC.

High power commercially available lasers typically emit light in the near infrared. As
diffraction limits the achievable focussed spot size to approximately the wavelength 
of the light, it is necessary to convert the laser light to a shorter wavelength to 
produce the small beam sizes required for a laserwire. However, light with a wavelength
shorter than 300~nm experiences strong absorption in most optical 
materials necessitating reflective focussing optics rather than transmissive ones. 
A laserwire at the Stanford Linear Collider (SLC) demonstrated micrometre-size 
scans with sub-micrometre resolution using an ultraviolet 
laser with a reflective focussing geometry~\cite{Alley1996,Ross2003}. This geometry 
prevents direct measurement of the laser focus and therefore calibration as well as 
imposing a very short scanning range due to the optical aberrations incurred when the focussing 
mirror is used off-axis - for example when either the laser or the mirror is moved to scan
the focussed laser spot across the electron beam. Transmissive optics afford
a much greater scanning range and allow direct measurement of the focussed spot size. 
Given these advantages, a compromise is to use a visible wavelength 
generated by frequency doubling a near infrared source, which increases the resolution
but reduces the peak power~\cite{Boogert2010}. 

Apart from the laserwire at SLC, several other laserwires have been demonstrated. A laserwire
at PETRA-III in DESY~\cite{Aumeyr2010} using a commercial Q-switched pulsed laser at 
20~Hz made laserwire scans of \mbox{$\sigma~\sim~30~\mu$m} on electron 
bunches in a ring acclerator with a precision of 5~\%. An installation at the 
damping ring of the 
Accelerator Test Facility 2 (ATF2) used a continuous-wave laser 
with cavity enhancement to measure an electron beam size of 9.8~$\mu$m~\cite{Sakai2001}. 
Here, the very low Compton-scattering rate from the low power laser is balanced by the megahertz 
revolution frequency of the electron bunches in the ring. In these cases, the low 
repetition rate and low average laser power respectively preclude fast intra-train 
scanning as would be required at the ILC. Another laserwire installation at the 
linear extraction 
line of ATF2 has shown micrometre-sized scans using a gigawatt peak power Q-switched 
pulsed laser at 3.25~Hz~\cite{Nevay2013}. Despite the high resolution and signal level at
this installation, the low repetition rate of the laser system also prevents intra-train 
scanning. In each case, the laser technology limits the usability of the laserwire as 
a diagnostic for a future linear collider. 

In this paper, we consider the requirements for a laserwire at the ILC and present the 
results of a fibre laser system developed for use there.  Its suitability is demonstrated 
through the calculation of the yield of Compton-scattered photons for such a system 
at the ILC. 

\section{ILC Laserwire Specifications}

The ILC is a proposed linear collider that is intended to collide electrons and positrons 
initially at 250~GeV and eventually up to 500~GeV. Such a machine is needed to make precision
studies of the recently discovered Higgs boson and to investigate potential 
new discoveries at the Large 
Hadron Collider at CERN and will require $\sim$20 laserwires in various 
locations as a primary accelerator diagnostic~\cite{ilc2013}. Laserwires will be of crucial 
importance in the beam delivery system (BDS) after the main linac, where precise 
measurements of the electron beam size of the order of a micrometre are necessary to
ensure that the electron and positron beams can be focussed to the required nanometre 
sizes at the interaction point. The parameters of the electron beam in the ILC BDS are 
shown in Table~\ref{tab:ilcparams}.

\begin{table}
\caption{\label{tab:ilcparams} Nominal ILC BDS electron beam parameters.}
\begin{tabular}{l c c c}
\hline \hline
Parameter                       & Symbol                & Value               & Units \\ \hline
Beam Energy                     & E                     & 250                 & GeV \\
Normalised horizontal emittance & $\epsilon_{x}^{\ast}$   & 10                 & $\mu$m~rad \\
Normalised vertical emittance   & $\epsilon_{y}^{\ast}$   & 40                 & nm~rad \\
Train repetition frequency      & $f_{train}$            & 5                   & Hz   \\ 
Number of bunches per train     & $N_{train}$            & 1312                &      \\
Bunch repetition frequency      & $f_{bunch}$            & 1.81                & MHz  \\
Bunch duration                  & $\sigma_{\tau e}$     & 1                   & ps   \\ 
Number of electrons per bunch   & $N_{e}$               & 2\ $\times$\ 10$^{10}$ &  \\ \hline \hline
\end{tabular}
\end{table}

The luminosity of the laser photon and electron collision is linearly 
proportional to both the number of photons 
and the number of electrons as well as the cross-section for the interaction, which in this
case is the Compton cross-section. \citeauthor{Agapov2007}~\cite{Agapov2007} have shown 
that given typical detector efficiencies and operational experience of a laserwire, 
a wavelength of 532~nm, laser pulse length of \mbox{$\sigma_{\tau}~=$~1~ps} 
and focussed to a size of \mbox{$\sigma_{l}~=~1~\mu$m}, 
that a laser peak power of approximately 10~MW will be required for each of the 
laserwires planned for use in the ILC. 
Currently available commercial laser systems capable of this peak power 
can do so only at kilohertz repetition rates and typically have a low overall efficiency. 
Such a system would therefore preclude intra-train scanning at megahertz repetition rates, 
and given the low electron bunch train repetition rate of 5~Hz, would take 
much longer to achieve the required measurement precision. In the case of a
future linear collider, the emittance measurement system may be used
often for continuous feedback and tuning purposes and a system utilising 
intra-train scanning would decrease the measurement time as well 
as increase the precision. 
Additionally, the spatial quality of such laser systems can be 
unsuitable for achieving the necessary diffraction-limited 
focussed spot sizes.  

A peak power of 10~MW with a pulse length of 1~ps corresponds to an energy 
of 10~$\mu$J per pulse. High energy ultrashort laser pulses in the visible part of the
spectrum are typically produced by frequency doubling a source at 
$\lambda\sim$1~$\mu$m, which can be achieved with an efficiency of 60~\%~\cite{Rothhardt2011}. 
Therefore, pulses of $\sim$17~$\mu$J in the near infrared 
are required at the megahertz repetition rate to allow intra-train scanning of the 
electron bunches. 

\section{Fibre Laser}

A technology with great potential for laserwires and other accelerator applications is 
that of optical fibre lasers, which has progressed rapidly in recent years, 
allowing the efficient generation of high energy ultrashort pulses~\cite{Richardson2010}. 
The waveguide structure of optical fibres means they have excellent spatial quality 
and high efficiency due to the long interaction lengths with the pump sources.  However, high 
intensities that arise in the small core diameters of single mode step-index fibres can 
induce optical nonlinearities that affect the spectral and temporal properties of 
the output pulses and limit the maximum useful output~\cite{Limpert2009}. 
One solution to this problem has been the development 
of photonic crystal fibres (PCFs) that have large mode areas with nearly single mode 
spatial output that allow high energy pulse extraction~\cite{Limpert2005}. 
In particular, rod-type PCFs have the largest core diameters that can be up to 
100~$\mu$m~\cite{Brooks2006} 
and have been shown to produce 1~mJ, 800~fs output pulses at 100~kHz~\cite{Roser2007}.
Rod-type PCFs cannot be connectorised like normal optical fibres, but they are 
desirable as their large area allows the highest peak powers to be achieved without
optical nonlinearities.
Fibre lasers have excellent conversion of pump to laser light and in addition, 
are pumped by highly efficient diode sources leading to very high overall wall plug 
efficiency. This, together with the excellent spatial quality and reliability of 
fibre lasers make them attractive for applications such as laserwire.

To investigate the suitability of a fibre laser for a laserwire, a system was 
developed consisting of a commercial chirped pulse amplification (CPA) fibre laser 
followed by further amplification in an 80~cm long Yb-doped rod-type PCF 
(NKT Photonics DC-200/85-Yb-ROD).
The commercial laser (Amplitude Syst\`{e}mes) consists of a Yb:KYW oscillator at 
1036~nm producing 19~nJ pulses at 51.92~MHz, locked to a Rb stabilised external radio 
frequency signal generator. The 500~fs oscillator pulses are stretched to $\sim$200~ps 
and then the repetition rate reduced to 6.49~MHz by an acousto-optic modulator. 
The pulses (now 6~nJ) are amplified in two stages in Yb-doped 25~$\mu$m core diameter fibre and 
then pass through an electro-optic modulator (EOM) for macropulse shaping, producing 
$\sim$1.5~$\mu$J output pulses in bursts as short as 500~ns, which are used as a seed for
the PCF.

\begin{figure}
\includegraphics[width=0.48\textwidth]{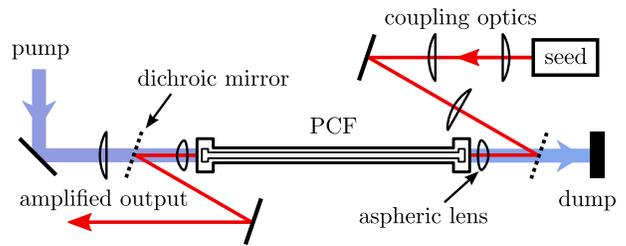}
\caption{\label{fig:pcfsetup} Experimental arrangement for amplification in a PCF with 
coupling and decoupling aspheric lenses as well as dichroic mirrors to separate the pump 
and seed beams.}
\end{figure}

The PCF has a core diameter of 85~$\mu$m to guide the seed, an inner cladding diameter of 
200~$\mu$m to guide the pump, and an outer cladding diameter of 1.7~mm. The 400~W, 
976~nm pump diode laser (Newport Spectra-Physics) is coupled into the PCF in a 
counter propagating geometry allowing independent optimisation of 
both the seed and the pump coupling as well as increased efficiency~\cite{Xue2007}. 
Fused silica endcaps 8~mm long and 10~mm diameter at both ends of the fibre 
allow the unguided light from the core and inner-cladding to expand before exiting 
through the glass air interface, reducing the possibility of damage to the 
fibre~\cite{Dawson2008}. The seed and pump are separated by dichroic mirrors at either 
end of the PCF, which is made entirely of silica with no polymer coating and is 
supported in a metal V-groove. Due to the high efficiency of fibre lasers and the 
large surface area to volume ratio, the fibre requires neither active nor passive 
cooling. The experimental arrangement is shown in Figure~\ref{fig:pcfsetup}.

\subsection{Burst Mode Amplification}

Whilst high peak power laser pulses are required at megahertz repetition rates to match 
the electron beam for intra-train scanning, the bunch trains have a low duty cycle 
(0.4~\%). The laser system was matched to produce pulses with a similar duty cycle, 
significantly reducing the average power and therefore cooling requirements as well
as avoiding wasting laser energy when there are no electron bunches. The 
EOM in the seed laser allows the necessary macropulse shaping and the pump output is 
modulated using the output of a signal generator as a gate to the power supply.

Simultaneously starting both the pump and the seed will not result in a uniform burst of 
amplified pulses as the pump has a finite rise time ($\sim$100~$\mu$s). 
With a simultaneous start, 
the PCF will not be pumped initially and the Yb dopant will absorb the incoming seed 
due to its quasi-three-level nature. During the rise time of the pump, the PCF becomes 
first transparent and then amplifies the seed. By turning on the pump before the seed by 
approximately the rise time of the pump a more uniform output can be achieved. However, 
by further increasing the delay of the seed burst with respect to the pump the Yb dopant 
is pumped without the seed depleting its excited upper-state population and energy can 
be stored in the PCF resulting in higher gain during the seed 
burst. After approximately the upper state lifetime of Yb in glass (0.8~ms), 
spontaneous emission of the excited population will reduce the available gain
and may lead to large laser pulses oscillating in the rod that can lead to
catastrophic damage~\cite{Limpert20052}.

\begin{figure}
\includegraphics[width=0.48\textwidth]{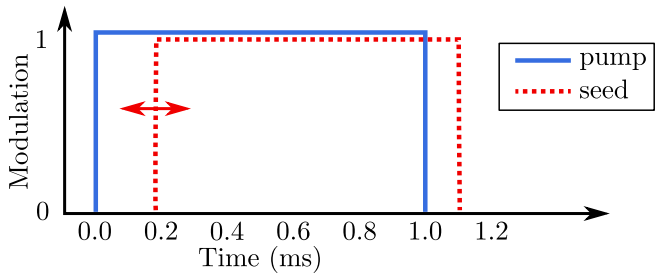}
\caption{\label{fig:tempstruct} Modulation envelopes of the seed and pump laser for 
burst operation with the pump preceding the seed for high gain.}
\end{figure}

An investigation was carried out to characterise this process and in doing so a method was 
demonstrated by which pulses almost an order of magnitude greater in energy than the 
steady-state can be achieved. The pump laser 
requires a minimum operation time of 1~ms, which corresponds to $\sim$6500~pulses of the seed.
Initially, the seed and pump bursts were operated simultaneously and then the start of the 
seed burst was delayed with respect to the start of the pump as shown in 
Figure~\ref{fig:tempstruct}. As the pump laser also has a finite fall time and some excited 
upper-state population still exists even when the pump is turned off, the seed burst was 
extended by 100~$\mu$s beyond the end of the pump to safely extract the stored 
energy. The system was operated at 2~Hz due to the limitation of the energy meter 
used to integrate the energy of the full burst and the system is capable of 
operating at higher burst repetition rates. 

As this technique relies on a build up of gain beyond the normal steady-state, 
it is to be expected that the high repetition rate seed pulses will deplete 
the upper-state population
before it is replenished by the pump. Consequently, the initial pulses in the burst
will experience a very high gain with subsequent pulses experiencing 
decreasing gain until the steady-state is reached. Although it is possible to compensate 
for this by modulating the incoming seed train before the PCF, this is 
still an active area of research, and can lead to
lower individual pulse energies in the burst~\cite{Breitkopf2012}. Despite the lack of 
uniformity in pulse energies, this scheme serves to investigate the potential
output from a fibre laser and could also be compensated for by pulse
energy normalisation in a laserwire system.

\subsection{Characterisation}

To characterise the output of the laser system, the pulse energies, spatial quality 
and temporal profiles were measured. The megahertz repetition rate of the pulses presents 
many practical challenges for conventional techniques used to measure these aspects of 
the pulses on an individual basis.  

\subsubsection{Pulse Energies}

\begin{figure}
\includegraphics[width=0.48\textwidth]{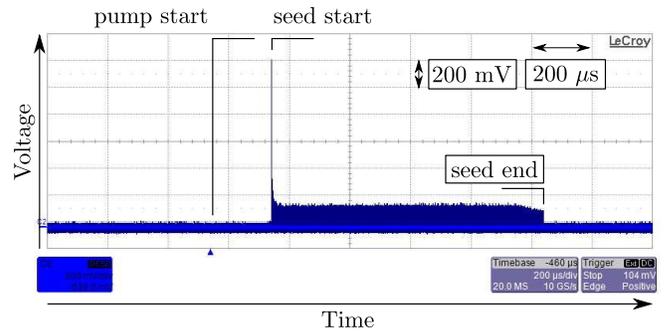}
\caption{\label{fig:pdtrace} Oscilloscope screenshot showing a 20 mega-sample photodiode 
trace of the amplified output exhibiting exponential decay of pulse energies 
to a steady-state level. The high number of pulses in the $\sim$1~ms burst 
results in a solid trace when viewed on this timescale and the aliasing of the
oscilloscope displays a much larger degree of noise than is actually present.}
\end{figure}

The amplified seed burst from the PCF with a seed delay relative to the pump of 0.13~ms 
is depicted in a photodiode trace in Figure~\ref{fig:pdtrace}, which shows a high 
initial series of pulses that rapidly decay to a steady-state level. As the pulses 
vary in energy significantly they cannot be assumed to be same and  must 
be measured individually. The megahertz repetition rate is beyond the ability 
of the fastest commercially available energy meters to discriminate individual pulses. 
Therefore, an energy meter capable of measuring 
the energy of the total burst (Gentec QE25-SM-LP) was used in combination 
with a digitised trace of a fast (\mbox{$<\,35$~ps} rise time) photodiode 
(Newport 818-BB-25 InGaAs) 
to resolve the individual pulse energies. 
As the peak of the photodiode pulses is 
proportional to the laser pulse energy, the digitised photodiode signal can be used 
to subdivide the total energy of the burst. The energy of a given pulse in the burst 
is therefore given by 

\begin{equation}
\label{eq:pdenergy}
E_{(i)} = \left( \frac{E_{total}}{\Sigma_{i} V_{peak(i)}} \right) V_{peak(i)}
\end{equation}

\noindent where $E_{(i)}$ and $V_{peak(i)}$ are the pulse energy and peak voltage of
$i$th pulse respectively, and $E_{total}$ is the total energy of the burst. An oscilloscope 
with a high bandwidth, sample rate and sample memory was used (LeCroy WaveRunner 402-MXi) to 
digitise the photodiode trace of the total burst with 12~bit precision at 
10~giga-samples~s$^{-1}$. Two Fresnel beam splitters were used with several neutral 
density filters to reduce the intensity of the pulses to within the measured linear range of 
the photodiode and 10 waveforms recorded for statistical purposes. 

For the most accurate measurement of the pulse energies in the seed burst, the photodiode 
traces were recorded using two different scalings on the oscilloscope; firstly with a 
scaling such that the high voltage peaks of the initial laser pulses were sampled properly; 
and secondly scaled so that the initial pulses were clipped but the 
majority of the burst was digitised with a higher precision. This method of using two 
combined sets of photodiode traces reduces the uncertainty in the calculated pulse 
energy from approximately 20~\% to 1~\%, which is due to the sum of the peak voltage 
of several thousand pulses. Additionally, the extension of the seed beyond
the pump was chosen to give a discrete end to the pulse train making the number
of pulses in the burst clearly identifiable. 

\begin{figure}
\includegraphics[width=0.48\textwidth]{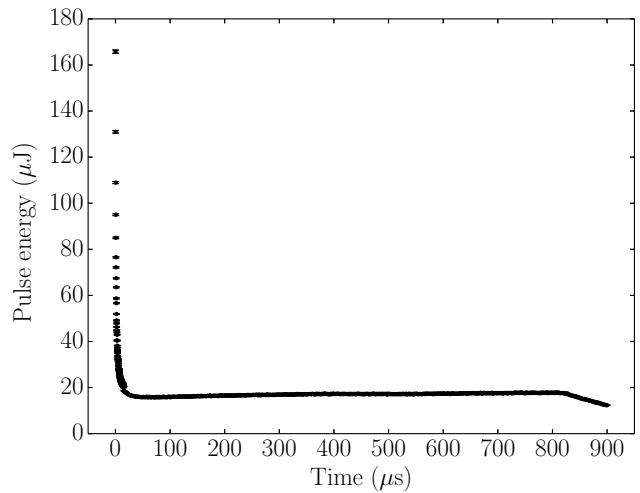}
\caption{\label{fig:energiesfull} Amplified pulse energies for a seed delay of 200~$\mu$s 
relative to the pump.}
\end{figure}

\begin{figure}
\includegraphics[width=0.48\textwidth]{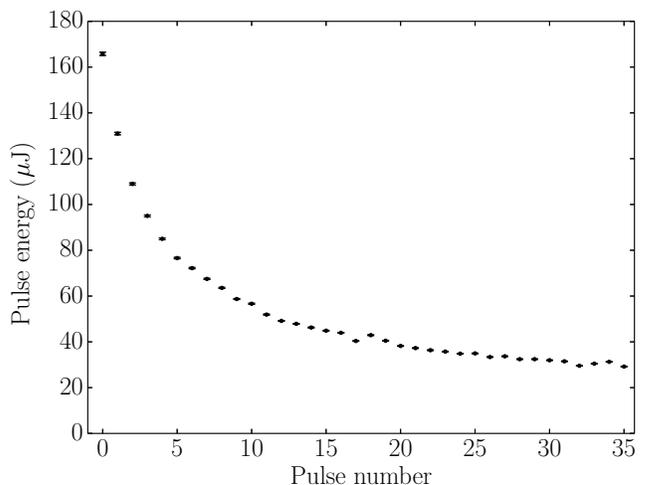}
\caption{\label{fig:energieszoom} Amplified pulse energies at the beginning of the seed 
burst shown in Figure~\ref{fig:energiesfull}.}
\end{figure}

The pump level was increased until sharp spikes appeared in the amplified spectrum - a sign 
of the presence of amplified spontaneous emission (ASE) and an indication that additional 
pump energy would not produce further useful amplified output. The incident 
pump burst energy was \mbox{297~$\pm$~1~mJ} and the input seed burst energy was 
6.43~$\pm$~0.02~mJ, 
which corresponds to an input pulse energy of 0.95~$\pm$~0.01~$\mu$J. 
A seed delay of 200~$\mu$s was found 
to give the maximum pulse energy of the first amplified pulse and in this 
case the total energy of the output 
amplified seed burst was 100.1~$\pm$~0.2~mJ. The output burst energy along with the 
digitised photodiode traces was used to calculate the pulse energies of the 5842 pulses 
shown in Figure~\ref{fig:energiesfull}. These show a maximum pulse energy of 
165.8~$\pm$~0.4~$\mu$J with an exponential decay 
in pulse energy to a steady-state. The mean pulse energy from pulse number 200 to 
5000 is 17.0~$\pm$~0.6~$\mu$J with a pulse to pulse variation of 0.5~\%. The linear 
decrease in pulse energy at the end of the pulse train is indicative of the pump turn 
off and depletion of gain in the PCF. The high initial pulse energies and following 
exponential decay are shown more clearly in Figure~\ref{fig:energieszoom}, which shows 
the first 35 pulses.

\begin{figure}
\includegraphics[width=0.48\textwidth]{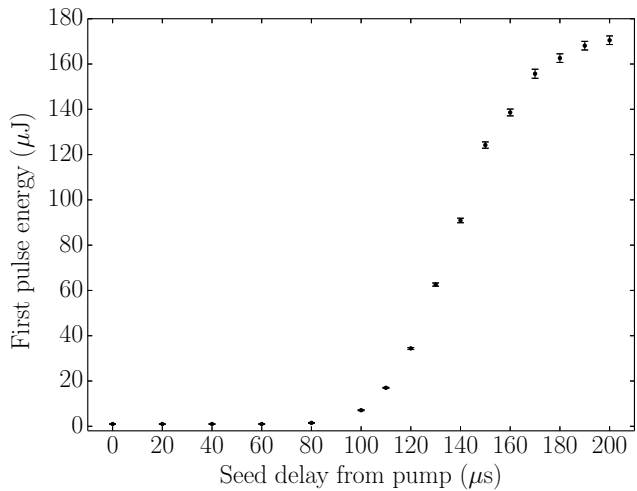}
\caption{\label{fig:energyvsdelay} Energy of the first pulse in the amplified train for 
various seed burst delays relative to the pump burst. }
\end{figure}

To investigate the relationship between the delay of the seed burst with respect to the 
pump pulse and the energy of the first pulse in the amplified seed burst, the seed 
delay was varied from 0 to 200~$\mu$s and the pulse energies of the full burst
measured. The energy of the first pulse in the burst as a function of seed burst delay 
is shown in Figure~\ref{fig:energyvsdelay}.

With the simultaneous start of the seed and pump, the energy of the first 
pulse in the output burst was 0.71~$\pm$~0.01~$\mu$J, which is lower than the incident pulse 
energy from the seed of 0.95~$\pm$~0.01~$\mu$J. At a seed delay of 200~$\mu$s, the gain of the 
initial pulse was 27.2~$\pm$~0.1~dB~m$^{-1}$.

The long term stability of the laser system was observed by making energy measurements 
over several periods of up to 12 hours, over which a variation of $<$~1\% was observed. 

\subsubsection{Spatial Quality}

\begin{figure}
\includegraphics[width=0.48\textwidth]{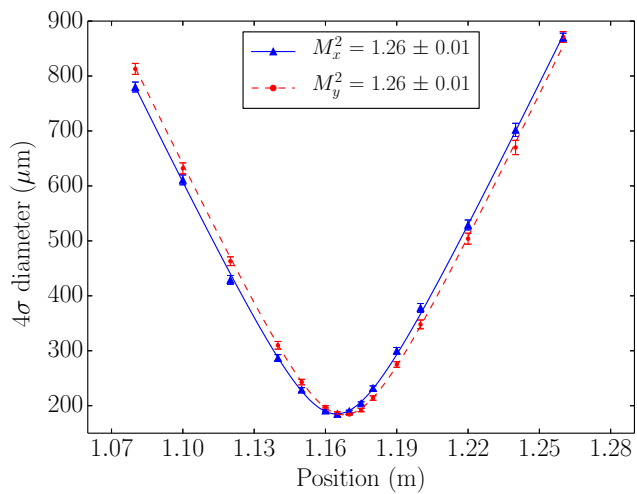}
\caption{\label{fig:msquare} $M^{2}$ measurement of amplified seed pulses with both the x 
and y fits shown.}
\end{figure}

The spatial quality was measured by focussing the maximum amplified output beam from the 
PCF using an \mbox{$f~=~$1~m} plano-convex lens. A laser beam profiler on a 
translation stage was used to 
measure the transverse profile at several points throughout the focus.  The 
4$\sigma$ diameter of the laser beam as it propagates can then be modelled by comparing it to a 
perfect Gaussian mode using the $M^{2}$ model~\cite{Johnston1998}. The $M^{2}$ parameter 
of the laser beam then linearly scales the focussed spot size for a given input beam size to
a lens and its focal length, with a perfect Gaussian by definition having an 
$M^{2}$ of 1. The 4$\sigma$ diameters and 
the $M^{2}$ model fits are shown in Figure~~\ref{fig:msquare} with measured 
values of \mbox{$M^{2}_{x}$~=~$M^{2}_{y}$~=~1.26~$\pm$~0.01}. This excellent 
spatial quality is well suited for creating diffraction limited focussed spot sizes. 

\subsubsection{Pulse Duration}

In a CPA laser system the pulses are temporally stretched before amplification to avoid 
optical nonlinearities as well as damage to the amplifier and compressed 
back to their original duration afterwards. The stretcher introduces linear group delay 
dispersion (second order phase) to the pulse, stretching it in time, and the compressor 
is designed to compensate for this as well as any higher order phase terms accumulated 
by the pulse during amplification. However, dispersion caused by nonlinear effects 
such as self-phase modulation cannot necessarily be corrected and in this case the 
compressed pulse will have a different temporal profile and duration to the initial 
pulse. Although high peak powers can be achieved in optical fibre amplifiers, 
the presence of nonlinearities limits the compression and therefore the final 
peak power attainable. It is therefore important to measure the compressed pulse duration.

\begin{figure}
\includegraphics[width=0.48\textwidth]{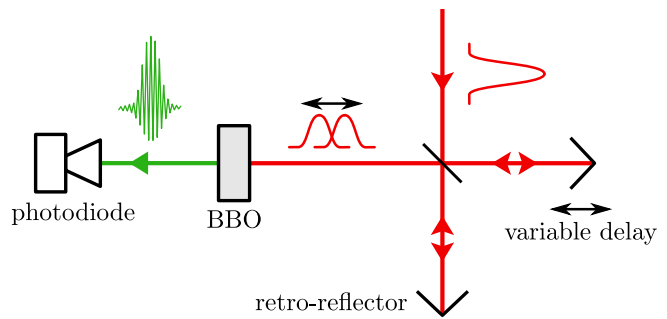}
\caption{\label{fig:autocorrelator} Schematic of an autocorrelator, consisting of an interferometer with beam splitter, retro-reflectors, a variable delay arm, BBO crystal and photodiode.}
\end{figure}

Ultrashort laser pulses well below 1~ns cannot be characterised by electronic devices 
alone as their response is too slow and therefore another technique must be used. 
Optical autocorrelation is the most common way of measuring these pulse 
durations~\cite{Monmayrant2010} and therefore an autocorrelator based on a scanning 
Michelson interferometer was used for our pulse duration measurement. This consists 
of a 50:50 
beam splitter to split the incoming pulse, two retro-reflectors to recombine 
the split pulses at the beam splitter and a beta-Barium Borate 
($\beta$-BaB$_{2}$O$_{4}$, BBO) frequency doubling crystal to analyse the 
interference between the two pulses as shown in Figure~\ref{fig:autocorrelator}. 
One arm of the interferometer is mounted on a 
translation stage and as it is scanned the delay between the two pulses 
is varied. The resultant frequency doubled output as detected by a photodiode 
shows the interference fringes, the envelope of which is the convolution of 
the pulse with itself. The standard technique involves scanning one arm of the
interferometer and continuously recording
data. In our case however, the low duty cycle of the bursts
and the significant time taken to digitise the waveforms makes this implementation 
unsuitable. Instead, waveforms were recorded at a series of static positions 
and the setup was mechanically 
agitated during the data recording, allowing the envelope to be observed at 
that position. A large number of waveforms were recorded at each position to 
ensure the full extent of the envelope was captured. To deconvolve the 
autocorrelation an assumption about the temporal profile of the pulse
is required. We assume a Gaussian temporal profile with a deconvolution
factor of $\sqrt{2}$.

\begin{figure}
\includegraphics[width=0.48\textwidth]{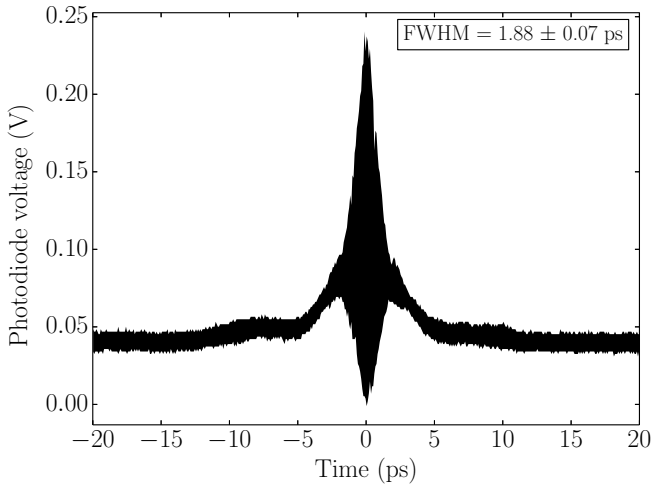}
\caption{\label{fig:ac:aslaser} Autocorrelation of the seed laser compressed without 
transmission through the final PCF amplification stage. }
\end{figure}

\begin{figure}
\includegraphics[width=0.48\textwidth]{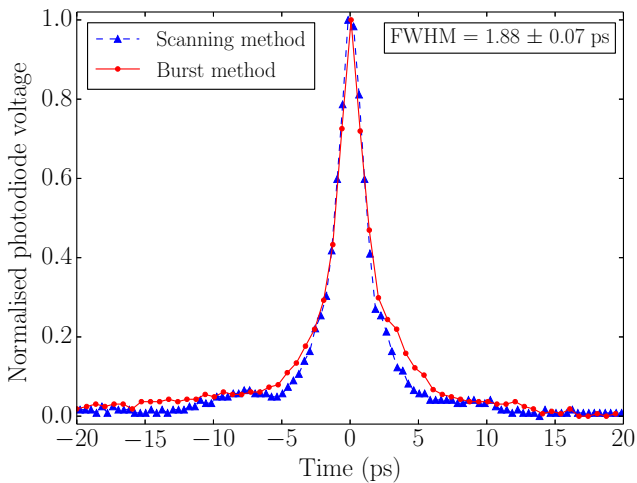}
\caption{\label{fig:accomparison} Envelope of autocorrelations of the seed laser made 
using both the conventional scanning method and the burst operation method showing good 
agreement with identical FWHM.}
\end{figure}

For verification, we compared pulse duration measurements made with both the method 
described above and also the conventional scanning autocorrelation technique. To 
perform this comparison, the EOM was turned off so that the laser output
consisted of a series of continuous pulses rather than operating in a burst mode. 
This enabled us to take data using both methods that could be directly compared. The 
conventional scanning autocorrelation is shown in Figure~\ref{fig:ac:aslaser}. 
The upper envelopes of this autocorrelation and that of the autocorrelation performed in 
burst mode are shown together in Figure~\ref{fig:accomparison} and show excellent agreement 
between both methods with a deconvolved full-width at half-maximum (FWHM) of 
1.88~$\pm$~0.07~ps in both cases. This is also indicative of the achievable pulse durations 
possible from this seed laser used with the PCF. 

\begin{figure}
\includegraphics[width=0.48\textwidth]{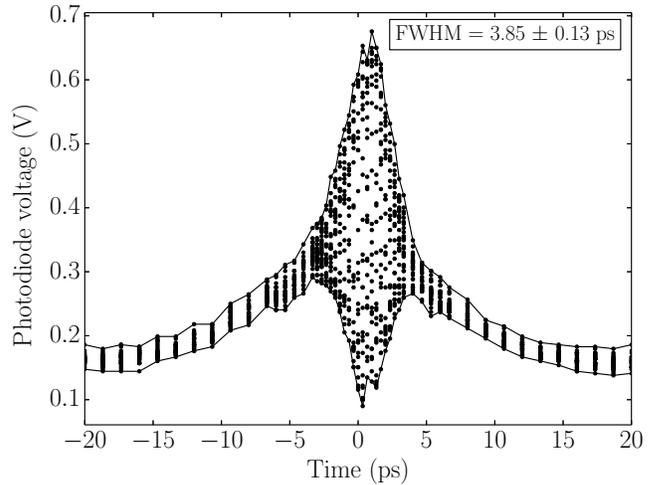}
\caption{\label{fig:ac:transmitted} Autocorrelation of the unamplified compressed seed 
pulses transmitted through the PCF.}
\end{figure}

Having measured the compressed output of the commercial system before the PCF, we then
measured the effect of transmission through the PCF both with and  without amplification. 
The autocorrelation of the unamplified compressed output 
from the PCF is shown in Figure~\ref{fig:ac:transmitted}, which shows a 
slightly wider peak with a deconvolved FWHM of 3.85~$\pm$~0.13~ps as well as 
slight wings. The transmission through the PCF introduces additional 
dispersion to the pulse that could not be fully compensated for by the 
compressor of the commercial laser used in this research, which has a fixed layout and a 
transmission efficiency of 75~\%. 

\begin{figure}
\includegraphics[width=0.48\textwidth]{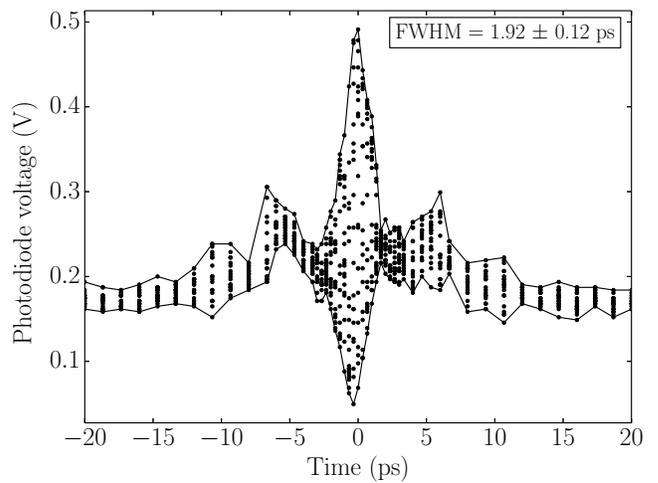}
\caption{\label{fig:ac:amplified} Autocorrelation of the first pulse in the amplified burst 
with an energy of 165.8~$\pm$~0.4~$\mu$J with the envelope shown.}
\end{figure}

\begin{figure}
\includegraphics[width=0.48\textwidth]{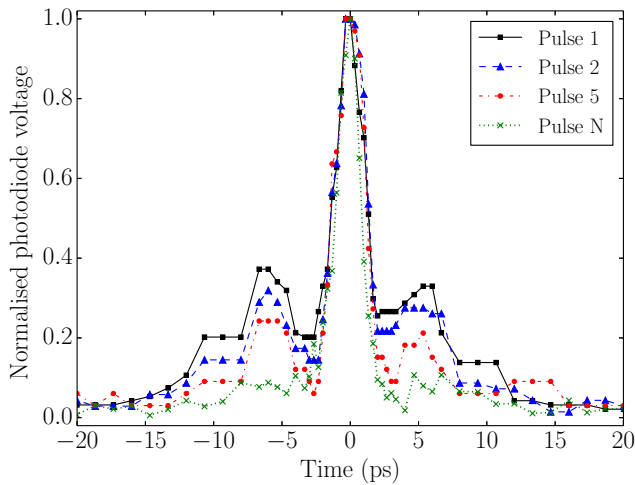}
\caption{\label{fig:ac:5pulse} Autocorrelation envelopes of pulses 1, 2, 5 and the $nth$ 
pulse, which is a pulse much later in the burst that is in the steady state. Each is 
normalised showing a clear reduction in the wings as the pulse 
energy decreases.}
\end{figure}

Figure~\ref{fig:ac:amplified} shows the autocorrelation of the first pulse in the 
amplified burst with the highest energy and therefore the most likely to have generated 
optical nonlinearities. In this case, only a small fraction of the amplified output 
from the PCF was compressed to avoid damage to the commercial compressor.
The autocorrelation shows a more complicated structure than the previous autocorrelations 
with a narrow peak with a FWHM of 1.92~$\pm$~0.12~ps as well as 
double wings on either side. Again, the fixed compressor arrangement prevents the 
possible compensation of additional dispersion introduced by the amplification. The 
wings in the autocorrelation indicate a temporal structure beyond a simple 
Gaussian such as a pre or post pulse~\cite{Trebino1990}. The 
central peak, as defined by the minima on either side, accounts for 40~$\pm$~1~\% 
of the total area of the 
autocorrelation and therefore that percentage of the total pulse energy is
contained within the central peak.
With the initial pulse energy of 
165.8~$\pm$~0.4~$\mu$J, and the compressor transmission of 75~\%, the peak power is 
correspondingly 25.8$~\pm$~1.7~MW. 
The normalised upper envelopes of the autocorrelations of this and subsequent pulses 
are shown in Figure~\ref{fig:ac:5pulse}. 

It can be seen that for pulses further into the train, the wings are smaller in the 
autocorrelations and a higher percentage of the pulse energy is contained within the 
central peak. As the pulses are also decreasing in energy, this would indicate 
an intensity dependence consistent with the presence of optical nonlinearities. 
The peak powers calculated from the percentage of energy in the central peak and the
pulse durations are shown 
in Table~\ref{tab:photons}. These peak powers would be significantly enhanced by 
the use of a compressor properly designed to compensate for the phase accumulated 
in the PCF amplifier.

\subsubsection{Polarisation}

As discussed in Section~\ref{sec:introduction}, the infrared laser source must
be frequency doubled to the visible part of the spectrum to achieve a smaller focussed
spot size. Frequency conversion requires that the input source be highly linearly polarised. 
Optical fibres do not preserve the polarisation of an input source due to inhomogeneities 
and stresses on the fibre caused by bending and torsion along its length. 
However, unlike most optical fibres the rod-type PCF is straight and therefore 
these effects were expected to be minimal.

The degree of linear polarisation (DOLP) was measured by recording the transmitted 
power through a polarising beam splitter as a function of angle and the maximum, 
$P_{max}$ and minimum, $P_{min}$ transmitted powers are used to determine the DOLP 
using Equation~\ref{eq:dolp}.

\begin{equation}
{\rm DOLP}\,(\%) = 100 \times \left(\frac{P_{max}-P_{min}}{P_{max}+P_{min}}\right)
\label{eq:dolp} 
\end{equation}

The DOLP of the input seed was measured to be 96.4~$\pm$~0.4~\% and the DOLP of the 
amplified output was 95.3~$\pm$~0.3~\%. This shows that the output is linearly 
polarised and that amplification in the PCF did not affect the polarisation despite 
the lack of specific polarisation maintaining structure. 

\subsubsection{Pointing Stability}

The laser system for a laserwire cannot not be placed next to the accelerator beam line
directly due to the damaging high radiation environment. Furthermore, this placement would 
preclude access to the laser system for maintenance.  
Therefore, the laser system must be located in a separate
area from the particle accelerator and the laser beam transported to the laserwire 
stations by either a system of mirrors (free-space transport) or in fibre.  With the 
required peak powers, it is not currently possible to transport laser pulses in fibre without
incurring optical nonlinearities that would distort the temporal and spectral properties of
the laser pulse.

Under free-space propagation, the pointing stability of the laser must be considered as the
angular jitter of the laser beam produces spatial jitter of the laser profile at the laserwire
lens. As many laser pulses are required to make a laserwire scan, the corresponding 
spatial jitter of the focussed laser beam is therefore convolved with the transverse 
size of the electron beam and increases the measured transverse size.

The pointing stability of the laser system was measured by measuring the laser profile 
centroid of the collimated output laser beam using a high resolution CCD beam profiler
at a distance of 3.518~m, where the laser beam size (4$\sigma$) was measured to be 
8.157~$\pm$~0.022~mm. The standard deviation of the centroid was meaured over a 20~min
period to be 3.820~$\mu$m, which corresponds to an angular jitter of 
1.09~$\pm$~0.03~$\mu$rad. For a laserwire system at a distance of 20~m~\cite{Nevay2014} 
producing a 1~$\mu$m focussed spot size using this laser beam, the spatial jitter at the
focus would be $<$\,10~nm.
When convolved with the laserwire scan this contribution would be negligible. At a greater
distance, the contribution may be non-negligible, but measurable and can be subtracted in 
analysis.

\begin{widetext}
\begin{table*}
\caption{\label{tab:photons} Summary of the measured properties and calculated peak 
powers of the pulses using the compressor transmission efficiency of 75~\%. The $nth$ 
pulse is representative of the mean of the steady-state pulses. The number of interacting
photons ($N_{\gamma~\mathrm{interacting}}$) was calculated by calculating the proportion of 
the total pulse energy within the full-width at half-maximum (FWHM) of the central peak. 
This was scaled assuming 75~\% compressor transmission and 60~\% frequency doubling efficiency.
The number of Compton-scattered photons is calculated from this assuming ILC electron beam
parameters detailed in Section~\ref{sec:application} for two possible laserwire collision 
angles of $\pi/2$ and $\pi/6$. }
\begin{tabular}{l c c c c c c c}
\\
\hline \hline
Pulse \#  & Energy           & FWHM            & Central Peak & Peak Power & 
$N_{\gamma~\mathrm{interacting}}$ & $N_{C}\,(\theta=\pi/2)$ & $N_{C}\,(\theta=\pi/6)$   \\ 
          & $\mu$J           & ps              &\%   & MW             & & & \\\hline
1         & 165.8~$\pm$~0.4  & 1.92~$\pm$~0.12 & 40  & 25.8~$\pm$~1.7 & $7.8\,\times\,10^{13}$ & 8496 & 22495\\
2         & 131.0~$\pm$~0.4  & 1.89~$\pm$~0.12 & 40  & 20.8~$\pm$~1.4 & $6.1\,\times\,10^{13}$ & 6755 & 17884\\
3         & 109.0~$\pm$~0.4  & 1.82~$\pm$~0.11 & 41  & 18.5~$\pm$~1.2 & $5.2\,\times\,10^{13}$ & 5845 & 15472\\
4         & ~95.1~$\pm$~0.3  & 1.72~$\pm$~0.11 & 45  & 18.6~$\pm$~1.3 & $5.0\,\times\,10^{13}$ & 5712 & 15118\\
5         & ~84.9~$\pm$~0.3  & 1.75~$\pm$~0.11 & 54  & 19.9~$\pm$~1.4 & $5.4\,\times\,10^{13}$ & 6082 & 16099\\ \hline 
$nth$     & ~17.0~$\pm$~0.1  & 1.73~$\pm$~0.11 & 100 & ~7.4~$\pm$~0.5 & $2.0\,\times\,10^{13}$ & 2265 & ~5994\\ 
\hline \hline
\end{tabular}
\end{table*}
\end{widetext}

\section{Application as a Laserwire\label{sec:application}}

In order to assess the suitability of the developed laser system for a laserwire, 
the yield of Compton-scattered photons from a laser pulse with an electron bunch 
must be considered along with the detector efficiency and required precision. Reviews of 
luminosity calculations and Compton-scattering cross-sections are given 
in~\cite{Suzuki1976,Tenenbaum1999} and a brief summary is given here. 

The number of Compton-scattered photons is given by the product of the luminosity 
$L\,(\theta)$ and the compton cross-section $\sigma_{C}\,(\omega)$, which are 
dependent on the collision angle~\cite{Klein1929,Miyahara2008}. The 
Compton cross-section is given by the product of the Thomson cross-section 
\mbox{($\sigma_{T}~=~0.665\,\times\,10^{-28}~$m$^{2}$)} with $f\,(\omega)$, which  is 
given by

\begin{eqnarray}
f\,(\omega) = \frac{3}{4} \left\{ \frac{1 + \omega}{\omega^{3}} \left[ \frac{2 \omega\,
(1 + \omega)}{1 + 2 \omega}
- \ln\,(1 + 2 \omega) \right] + \right.\nonumber 
\\[3ex]
\left. \frac{\ln\,(1 + 2 \omega)}{2 \omega} - \frac{1 + 3 \omega}{(1 + 2 \omega)^2}\right\}
\label{eq:comptonf}
\end{eqnarray}

\noindent where $\omega$ is the normalised energy of the laser photons in the electron rest 
frame, is given by

\begin{equation}
\omega = \frac{h E_{b}}{\lambda c^{3} m_{e}^2}\,(1+\cos \theta)
\label{eq:normedenergy}
\end{equation}

Here, $\lambda$ is the laser wavelength and $E_b$ the electron beam energy ($h$ is 
the Planck constant, $c$ the speed of light and $m_e$ the electron mass). The luminosity 
$L\,(\theta)$ can be decomposed into the product of the luminosity 
for a head on collision $L\,(0)$ and a reduction factor $R\,(\theta)$

\begin{eqnarray}
L\,(0) = \frac{n_{\gamma}\,n_{e}}{2 \pi\,\sqrt{\sigma_{\gamma x}^{2} + \sigma_{e x}^{2}} 
\,\sqrt{\sigma_{\gamma y}^{2} + \sigma_{e y}^{2}}} \label{eq:headonluminosity}
\\
R\,(\theta) = \frac{1}{\sqrt{1 + \left(\frac{\sigma_{\gamma z}^{2} + \sigma_{e z}^{2}}
{\sigma_{\gamma x}^{2} + \sigma_{e x}^{2}}\right)
 \tan^{2}{(\theta/2)}}} \label{eq:reductionfactor}
\end{eqnarray}

where $n_\gamma$ and $n_e$  are the number of photons in the laser pulse and electrons in a bunch 
respectively. The laser pulse and electron bunch are assumed to have Gaussian 
distributions with the corresponding sigma parameters in three orthogonal dimensions; 
$\sigma_{\gamma x}$, $\sigma_{\gamma y}$, $\sigma_{\gamma z}$ and $\sigma_{e x}$, $\sigma_{e y}$, 
$\sigma_{e z}$ 
respectively. For simplicity, the laser divergence is neglected, i.e. it is assumed that
the Rayleigh range of the laser is large compared to $\sigma_{e x}$.

Taking the ILC as an example, representative parameters of the electron beam in the 
BDS part of the accelerator where a laserwire is likely to be used are \mbox{$E_{b}$~=~250~GeV}, 
\mbox{$n_e$~=~2~$\times$~10$^{10}$}, \mbox{$\sigma_{e x}$~=~10~$\mu$m}, 
\mbox{$\sigma_{e y}$~=~1~$\mu$m} and $\sigma_{e z}$~=~300~$\mu$m~\cite{Agapov2007}. 
Assuming 75~\% transmission through a laser compressor, 60~\% frequency-doubling 
conversion efficiency (\mbox{$\lambda~=~$518~nm}), the measured pulse energies from
the laser system described are used to estimate the number of Compton-scattered 
photons for collision angles of both $\pi/2$ and $\pi/6$ as summarised in 
Table~\ref{tab:photons}. Only the pulse energy within the central peak 
of the laser pulse is considered.

The collision angle of $\pi/6$ in the horizontal plane increases the 
luminosity but effectively reduces the distance over which the laser beam
diverges across the electron beam. With the electron beam sizes 
mentioned, the reduced Rayleigh range is still over double the width of the 
electron beam. Even when present, such effects do not preclude the use of a 
laserwire as has been recently shown~\cite{Nevay2014}.

Assuming a detector efficiency of 5~\%~\cite{Agapov2007}, the number of 
detected Compton-scattered photons from the highest energy pulse with
an angle of $\pi/2$ would be 425, which is enough for a good electron beam emittance 
measurement~\cite{Agapov2007}. With an angle of $\pi/6$, this is increased
to 1125 detected photons, which would significantly improve the detector statistical
accuracy. Using the steady state pulse energy at this angle, 300 photons would be 
detected, which is still adequate for a laserwire at the ILC.

\section{Conclusions \& Outlook}

A fibre laser system suitable for an ILC laserwire has 
been demonstrated. Using a novel burst mode amplification regime, pulses an 
order of magnitude greater in energy than achievable in the steady-state have been shown 
with a maximum pulse energy of \mbox{165.8~$\pm$~0.4~$\mu$J}. 
This method was shown to be controllable by varying the relative timing of the pump and the
seed bursts allowing a range of initial pulse energies. The methods developed to 
characterise the 6.49~MHz pulses were detailed, which allowed
measurement of the pulse energy and autocorrelation on a pulse by pulse basis. The 
spatial quality of the amplified laser output was measured and shown to be excellent,
which would allow small focussed spot sizes to be achieved. 

The demonstrated high peak powers could be further improved with 
a specifically designed compressor that properly compensates for the additional group
delay dispersion introduced in the PCF to produce shorter duration pulses and 
consequently higher peak powers. In this case, the useful pulse energy within the pulse 
duration would be significantly increased from 40~\% demonstrated in the highest energy 
pulse. Furthermore, 
stretching the oscillator pulses to approximately 1~ns is possible and would significantly 
reduce the nonlinearities accumulated. Research on better compression and more efficient 
gain extraction in the PCF is underway.

Regarding the use of the laser for a laserwire diagnostic, the system described would 
currently be suitable for use with the ILC and if required, averaging of multiple 
electron bunches would decrease the uncertainty of the measurement. The demonstrated
laser pulse train varies significantly in pulse energy with an initial pulse 
approximately an order of magnitude greater than the steady state. Normalisation could 
be used to account for this variation as the pattern was found to be very stable. Whilst
the steady state pulse energies are sufficient, it may be highly desirable to have a 
different operational mode for a laserwire where a small number of high energy pulses
can be generated allowing alignment or calibration.

The background conditions 
for such an accelerator have yet to be accurately simulated
and this is an active area of research. With this information, the required laser pulse
energy could be more precisely defined. Simulation of the laserwire signal in 
combination with the accelerator backgrounds is also highly relevant as although
laserwire diagnostics have been demonstrated at test facilities, the high energy case
of the ILC will provide multi-GeV Compton-scattered photons that may be more 
distinguishable from background sources than those at the lower energy test facilities. 
In addition, improvement of the detector efficiency would have a significant impact on 
the required laser peak power.

This system demonstrates the peak powers required for a laserwire diagnostic as well as 
excellent spatial quality, which in addition to the high electrical efficiency of fibre 
lasers demonstrates their suitability for accelerator applications.

\begin{acknowledgments}
The research leading to these results has received funding from the Science and 
Technology Facilities Council via the John Adams Institute, University of Oxford and  
CERN.
\end{acknowledgments}

\bibliography{nevay2013fibre}

\end{document}